%% file: I05345I.tex
\documentclass[useAMS,usenatbib,usegraphicx]{mn2e}

\usepackage{latexsym}
\usepackage{amsmath}
\usepackage{amsfonts}
\usepackage{amssymb}
\usepackage{makeidx}
\usepackage{graphicx}
\usepackage{natbib}

\newcommand{\um}{\mbox{\,$\mu$m}}

\title[Massive Star-formation around IRAS\,05345+3157\\I: The Dense Gas]{Massive Star-formation around IRAS\,05345+3157\\I: The Dense Gas}
\author[Katherine I.\ Lee, Leslie W.\ Looney, Randolf Klein and Shiya Wang]{Katherine I.\ Lee$^{1}$\thanks{E-mail:
ijlee9@illinois.edu}, Leslie W.\ Looney$^{1}$, Randolf Klein$^{2,3}$ and Shiya Wang$^{4}$ 
\\
$^{1}$Department of Astronomy, University of Illinois at Urbana-Champaign, 1002 W Green St, Urbana 61801, USA\\
$^{2}$SOFIA-USRA, NASA Ames Research Center, Mail Stop N211-3, Moffett Field, CA 94035, USA \\
$^{3}$Department of Physics, University of California at Berkeley, 366 Le Conte Hall, Berkeley, CA 94720, USA \\
$^{4}$Department of Astronomy, University of Michigan at Ann Arbor, 500 Church St., Ann Arbor, MI 48109, USA
}
\begin{document}

\date{}

\pagerange{\pageref{firstpage}--\pageref{lastpage}} \pubyear{2011}

\maketitle

\label{firstpage}

\begin{abstract}
We present observations of the intermediate to
massive star-forming region I05345+3157 using the molecular line tracer CS($2-1$) 
with CARMA to reveal
the properties of the dense gas cores.
  Seven gas cores are
  identified in the integrated intensity map of CS($2-1$).  Among these,
  core 1 and core 3 have counterparts in the $\lambda$~=~2.7~mm continuum data.
  We suggest that core 1 and core 3 are star-forming
  cores that may already or will very soon harbor
  young massive protostars.
  The total masses of core 1 estimated from the LTE method and dust emission by assuming a gas-to-dust ratio  
  are 5$\pm$1 M$_{\sun}$ and 18$\pm$6 
  M$_{\sun}$, and that of core 3 are 15$\pm$7 M$_{\sun}$ and 11$\pm$3 M$_{\sun}$.  
  The spectrum of core 3 shows
  blue-skewed self-absorption, which
  suggests gas infall -- a collapsing core.
  The observed broad linewidths of the seven gas cores indicate non-thermal motions.
  These non-thermal motions can be interactions with nearby outflows or due to the initial turbulence;
  the former is observed, while the role of initial turbulence is less certain.  
  Finally, the virial masses of the gas cores are larger than the LTE masses, which for a bound core 
  implies a requirement on the external pressure of $\sim 10^{8}$ K cm$^{-3}$.
  The cores have the potential to further form massive stars.   
 
\end{abstract}


\input{introI}

\input{observationsI}
\input{resultsI}

\input{discussionI}
\input{conclusion}

\section{Acknowledgments}
We thank the anonymous referee for the valuable comments.  
We acknowledge support from the
Laboratory for Astronomical Imaging at the University of Illinois.
We thank the OVRO/CARMA staff and
the CARMA observers for their assistance in obtaining the data.
Support for CARMA construction was derived from the states
of Illinois, California, and Maryland, the Gordon and Betty
Moore Foundation, the Eileen and Kenneth Norris Foundation,
the Caltech Associates, and the National Science Foundation.
Ongoing CARMA development and operations are supported by
the National Science Foundation under cooperative agreement
AST-0540459, and by the CARMA partner universities.

\bibliographystyle{mn2e}
\bibliography{I05345}
\clearpage

\end{document}

%% file: introI.tex
\section{INTRODUCTION}
\begin{figure*}
  \begin{center}
    \includegraphics[scale=0.5,angle=-90]{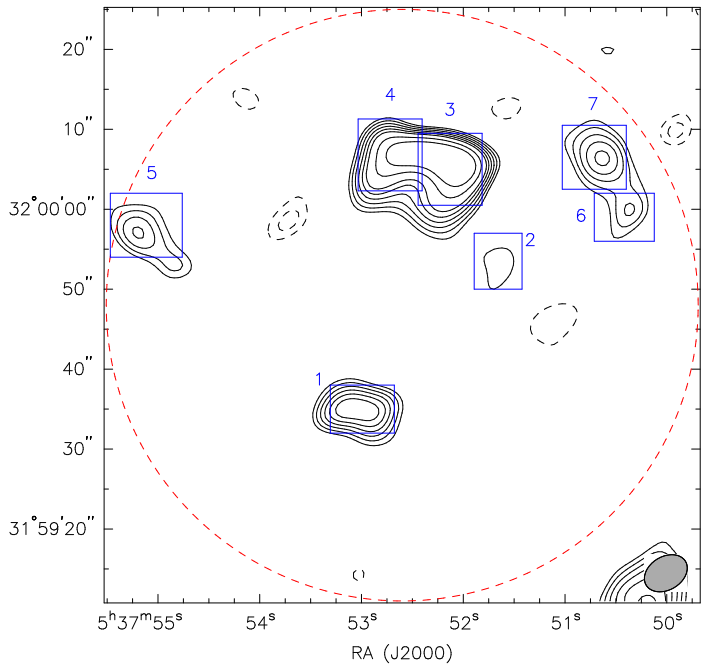}
      \caption{Contour map of CS $J=2-1$ integrated intensity with the FWHM beam
      size shown in the lower right corner.  The dashed circle
      is the size of the primary beam of the 10-meter dishes.  The synthesized beam size,
      shown in the right-bottom corner, is 4.3\arcsec$\times$5.9\arcsec with
      P.A.=-57.5\degr.  The noise is 0.25 Jy beam$^{-1}$ km s$^{-1}$, and the
      contours are 5$\sigma$, 6$\sigma$, 7$\sigma$, 8$\sigma$, 9$\sigma$, 10$\sigma$, 12$\sigma$, 14$\sigma$, positive and negative.
      The boxes are drawn based on the 3$\sigma$ detections and are used to define cores.}
    \label{fig:csmap}
  \end{center}
\end{figure*}

Despite the well-known fact that most stars form in clusters
\citep[e.g.,][]{2003ARA&A..41...57L}, little is known about the detailed formation
process of massive stars or star clusters.  In contrast to the well-studied
mechanisms for low mass star formation, the difficulty for massive star
formation is mainly due to a massive star's high radiative power.  The strong
radiation hinders infalling material in the envelope from further accreting
onto the central source.  Two major theories have been proposed to solve the
puzzle.  Accretion can still be accomplished through a high accretion rate
generated by the turbulent environment \citep{2003ApJ...585..850M} and also
through disks \citep[e.g.][]{2005IAUS..227..231K} with outflows that greatly
reduce the radiative pressure \citep{2005ApJ...618L..33K}.  On the other hand,
massive stars could also form from competitive accretion or stellar mergers
\citep{2001MNRAS.323..785B,2004MNRAS.349..735B} in high stellar density
regimes.  The formation of star clusters can be due to fragmentation
of molecular clouds \citep[e.g.,][]{2005MNRAS.356.1201B}, although
\citet{2007ApJ...656..959K} concluded that the fragmentation of pre-stellar massive 
cores can be avoided through heating, allowing the formation of a massive star from 
one massive core.  
Only through a detailed investigation of the
initial conditions in a protocluster system can we better address the problem
of massive star formation.

Massive protostars have strong interactions with their ambient
environment.  They disperse their natal clouds through
stellar winds and UV radiation \citep[see the review by][]{2007ARA&A..45..481Z}. 
After such dynamical interactions, the
ambient environment changes enormously and no longer preserves the initial
conditions from when massive protostars were first born.  Therefore, in order to
examine the very early conditions of the onset of massive star formation,
identifying stellar objects at early evolutionary stages \citep[e.g.,][]{2009ApJ...699..150S} in protoclusters
appears to be very essential.  \citet{2007prpl.conf..165B} suggested that the
early evolution has four stages starting with High-Mass Starless Cores
(HMSCs),  which then accrete low and intermediate mass protostars, leading to
High-Mass Protostellar Objects (HMPOs) and finally the (still deeply embedded)
final star.  

Here we report on the massive star-forming region around the FIR-bright
IRAS source 05345+3157, hereafter I05345.  The
IRAS source is associated with an IR cluster surrounded by molecular gas at a
distance of 1.8 kpc \citep{1992A&AS...93..525H}. HCO$^{+}$(2-1) emission east
of I05345 presents complex and clumpy structures, suggesting that the source is on the verge of
collapse through the contraction of local density peaks
\citep{2002ApJ...570..758M}. A massive outflow has been observed in CO(2-1)
\citep{2005ApJ...625..864Z,2009A&A...499..233F}, a typical signature of young stellar objects.
\citet{2005ApJS..161..361K} shows a ring of dust emission around the cluster I05345
in 850\,$\micron$ continuum data. Two compact and massive dust clumps
are located in the ring to the east of the cluster: I05345\,\#1 and
\#2, with the mass of 48 M$_{\sun}$ and 38 M$_{\sun}$, respectively.  
Futhermore, \citet{2008A&A...477L..45F} and \citet{2009A&A...499..233F} concluded 
several objects at different evolutionary stages in this region with SMA and PdBI observations, 
including prestellar core candidates, young intermediate to high mass class 0 protostars, 
and an early-B ZAMS star.  The same papers also revealed two condensations in 
N$_{2}$D$^{+}$(2-1) and the nature of the deuterated cores have been studied in detail.  

The previous observations showed 
the possibility of active star formation in I05345.
The main goal of this paper is to study star formation by investigating
the dense gas in the I05345 environment.  Our strategy uses the CS($2-1$) transition as a
probe of gas cores to comprehensively study gas
properties.  We use the D array configuration of the
Combined Array for Research in Millimeter-wave Astronomy (CARMA) to
resolve-out large-scale structures and trace dense gas fragments ($\sim$
10$^{5}$ cm$^{-3}$) inside the cloud.  In this paper we present $\sim$5\arcsec\ angular
resolution data of CS($2-1$) and 2.7 mm continuum.  In the companion 
paper by \citet{klein2011} (hereafter paper II), we examine the infrared properties of 
the fragments.  With the two papers, we obtain a general picture of star formation 
in I05345.

%% file: observationsI.tex
\section{OBSERVATIONS}
\subsection{CARMA observations}

Observations of the line transition CS $J=2\rightarrow1$ ($\nu = 97.981$ GHz)
toward the clumps I05345 \#1 and \#2 \citep{2005ApJS..161..361K} 
were made with CARMA in 2008 June.  The observations used CARMA 
D array at $\lambda=2.7$ mm. The CS($2-1$) transition was placed in a
correlator window of 8 MHz width and a velocity resolution of 0.37 km
s$^{-1}$.  Two 500 MHz bands were set up for continuum observations.  The
phase center was chosen between I05345 \#1 and \#2 at
$RA=05^{\text{h}}37^{\text{m}}52.6^{\text{s}}$ and $DEC=+31^{\circ}59\arcmin48.0\arcsec$ (J2000) using the local standard of
rest velocity V$_{\text{LSR}}$ of $-18.6$\,km\,s$^{-1}$
\citep{1996A&A...306..267S}.  0555+398 was used as the phase calibrator.
3C454.3 and Uranus were used for the bandpass calibration and the flux
calibration, respectively.  The system temperature ranged from 50 K to 300 K
with 200 K (single sideband) being a representative value.  The baselines
ranged from 11 meters to 150 meters.  The channel maps were generated with
natural weighting, producing a synthesized beam of $5.7\arcsec
\times 4.2\arcsec$.  
The primary beam size of the CARMA combined array is 85\arcsec at 98 GHz.  
The data reduction was done using the MIRIAD package
\citep{1995ASPC...77..433S}.
The amplitude calibration is estimated to be 10\%, and the
uncertainties discussed afterwards are only statistical not systematic.

%% file: resultsI.tex
\section{RESULTS}
\subsection{CS Zeroth Moment Map and Spectra}
\label{sec:CSobs}

\begin{table*}
 \centering
 \begin{minipage}{140mm}
  \caption{Parameters of the identified cores}
  \label{tbl:parameters}
  \begin{tabular}{@{}cccccccccc@{}}
  \hline
   Label & R.A. & Dec. & $a_{0}$ & $b_{0}$ & P.A.$_{0}$ & a & b & P.A. & R \\
        & (J2000.0) & (J2000.0) & (arcsec) & (arcsec) & (deg) & (arcsec) & (arcsec) & (deg) & (10$^{-2}$ pc) \\
 \hline
 Core 1 & 05:37:53.03 & 31:59:34.9 & $12.0\pm 1.6$ & $6.9\pm 0.6$ & $+83.5\pm 5.3$ & $10.9\pm 1.8$ & $4.9\pm 0.8$ & 78.8 & $3.2\pm 0.4$ \\
 Core 2 & 05:37:51.67 & 31:59:53.0 & $ 8.8\pm 2.2$ & $7.5\pm 1.7$ & $-25.4\pm 55.6$ & $ 7.2\pm 2.4$ & $5.5\pm 2.3$ & -3.7 & $2.7\pm 0.7$ \\
 Core 3 & 05:37:52.19 & 32:00:05.4 & $12.3\pm 3.9$ & $8.9\pm 1.5$ & $+63.9\pm 27.3$ & $11.4\pm 4.2$ & $7.0\pm 1.9$ & 58.8 & $3.9\pm 0.9$ \\
 Core 4 & 05:37:52.58 & 32:00:06.5 & $11.8\pm 4.0$ & $7.4\pm 1.2$ & $-83.6\pm 14.9$ & $10.5\pm 4.5$ & $5.9\pm 1.5$ & -88.3 & $3.4\pm 0.9$ \\
 Core 5 & 05:37:55.14 & 31:59:57.0 & $12.3\pm 2.6$ & $8.1\pm 1.0$ & $+48.2\pm 17.2$ & $11.5\pm 2.8$ & $5.9\pm 1.4$ & 45.6 & $3.6\pm 0.6$ \\
 Core 6 & 05:37:50.39 & 32:00:01.0 & $11.8\pm 7.9$ & $6.8\pm 1.3$ & $-14.2\pm 20.7$ & $10.7\pm 8.7$ & $4.6\pm 1.9$ & -9.3 & $3.1\pm 1.4$ \\
 Core 7 & 05:37:50.64 & 32:00:06.3 & $11.1\pm 0.9$ & $7.7\pm 0.4$ & $+37.2\pm 7.7$ & $10.3\pm 1.0$ & $5.1\pm 0.6$ & 36.0 & $3.2\pm 0.2$ \\
 \hline
 \end{tabular}
\medskip
 a$_{0}$, b$_{0}$ and P.A.$_{0}$ are the observed major axis, minor axis and position angle of the cores.  a, b and P.A. are the deconvolved major axis, minor axis and position angle.  R is the equivalent radius, $Sqrt{ab}/2$.
 \end{minipage}
 \end{table*}

Figure \ref{fig:csmap} shows the velocity-integrated intensity map of
CS($2-1$); the dashed circle is approximately the primary beam size of
the 10-m antennas in CARMA ($\sim 69 \arcsec$) at $\lambda$=2.7 mm.  We
identify seven cores 
with peak intensities above $5\sigma$ (1 $\sigma$
$\sim$ 0.25 Jy beam$^{\text{-1}}$ km s$^{\text{-1}}$) inside the primary beam of the integrated CS map .  
Due to the interferometric effect that introduces the negative components in the map, 
it is possible that the weak emission associated with core 2 is spurious. 
The boxes shown in the right panel of Figure \ref{fig:csmap} are used for deriving 
the parameters of the cores (see the paragraph below) and are drawn based on the 3$\sigma$ detection.

For this study, we
only focus on the cores inside the primary beam size of the 10-m OVRO dishes
and neglect the emission outside the primary beam, which has a lower sensitivity
and signal-to-noise ratio.  Core 1 is isolated in the south.  Cores 3 and 4
are spatially close and entangled.
In Table~\ref{tbl:parameters} the parameters of each core are specified,  
including the observed sizes (major axis a$_{0}$, minor axis b$_{0}$ and
position angle P.A.$_{0}$), the deconvolved sizes (major axis a, minor axis b
and position angle P.A.) and the equivalent radius R=$\sqrt{ab} /2$.  The
major and minor axes are derived by fitting a circular or elliptical Gaussian
profile in the velocity-integrated map of CS.  At the distance of 1.8 kpc, the
sizes of these cores range from 0.05 pc to 0.08 pc (see Table
\ref{tbl:parameters}),  which is
the typical size of transition groups from Hot Molecular Cores (HMCs) to
Ultracompact HII regions (UCHIIs) \citep{2007prpl.conf..165B} or of 
low-mass envelopes \citep[e.g.,][]{2000ApJ...529..477L}.

\begin{figure*}
\begin{center}
   \includegraphics[scale=0.5,angle=-90]{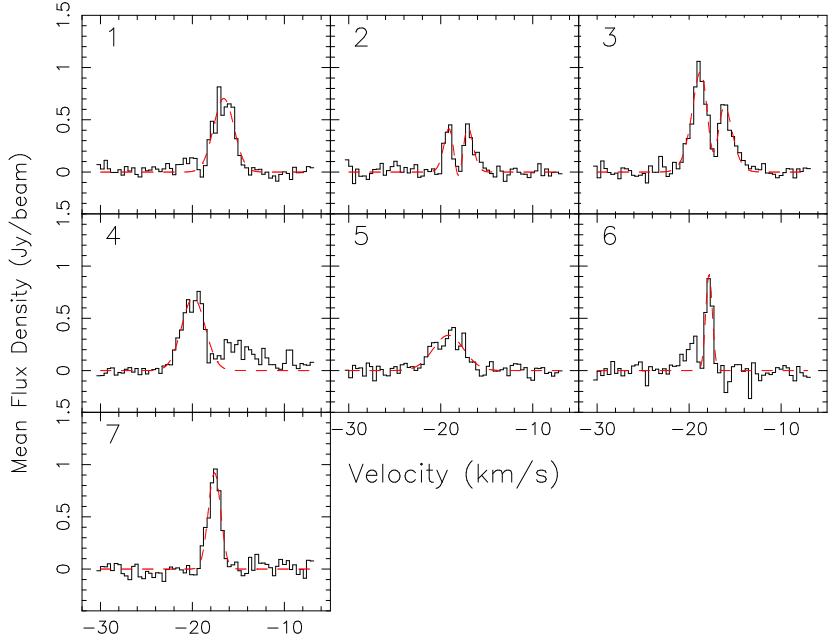}
   \caption{Spectra of identified CS cores.  V$_{\small LSR}$ is -18.6 km/s. The dashed lines
are the best fit Gaussians.}
   \label{specfit}
\end{center}
\end{figure*}

The spectra of the seven cores averaged over the boxes in
Figure~\ref{fig:csmap} are shown in Figure \ref{specfit}. 
Most of the spectra show a strong single-peaked emission, while
the spectra of core 2 and 3 are double peaked. 
The spectra are fitted by
Gaussian profiles as shown in Figure \ref{specfit}.  Since the double peaks in
the spectra of core 2 and core 3 may result from self-absorption
(Sect.~\ref{ssec:cs}), we fit the 
profiles with one Gaussian emission and one Gaussian absorption line
superimposed.  Table \ref{tbl:fits} lists the fitting parameters 
for each core (amplitude, peak position of velocity, FWHM).  
Note that the spectrum of core 4 shows a red wing emission
and that of clump 6 shows a blue wing emission above the fitted line.  
The line wings may be due to the outflow driven by core 3 (see Sect.~\ref{sec:dynamics}).

\subsection{Millimeter Continuum Data}
\label{sec:MMobs}

Figure \ref{fig:cont} shows the contours of the CS intensity map overlaid on
the 2.7\,mm continuum data from the same observation.  There are three 2.7 mm
continuum emission peaks in the map (MM1, MM2 and MM3).  MM1 in the south is
the brightest core ($22.7\pm 7.2$ mJy). 
MM2, the faintest ($6.0\pm 1.4$ mJy), is in the north accompanied by the brighter 
MM3 ($13.8\pm 3.6$ mJy).  These sources are all unresolved.

Comparing the $\lambda$=2.7 mm observation with CS cores, we see that core 1
is associated with MM1 although the two peaks are shifted by about 3$\arcsec$,
smaller than the beam size.
Core 3 coincides with MM3, while MM2 shows no CS emission (see below).
There are no other 2.7 mm emission features associated with the rest of the CS
cores.

MM3 and MM1 correspond to the clumps I05345 \#1 and \#2 for which
\citet{2005ApJS..161..361K}  derived masses of 48 M$_{\sun}$ and 38 
M$_{\sun}$, respectively.  Cores 2, 3 and 4 are all associated with the
northern clump I05345 \#1.  
In addition, the three cores C1, C2, and C3 in 225 GHz continuum observation
by \citet{2008A&A...477L..45F} coincide with  MM3, MM2, and MM1, respectively.

The cores MM1 and MM3 have been consistently detected in the
(sub-)millimeter continuum and also in the FIR by Spitzer (see paper II for details). 
Figure~\ref{fig:cont} shows the Spitzer 24 $\micron$ image; the bright northern source 
corresponds to our core 3 and the southern source corresponds to core 1.  
It is peculiar that at 2.7\,mm the northern source is the weaker one of the two
while at all other wavelengths from 24\,{\micron} to 1.3\,mm (the 225 GHz observation in \cite{2008A&A...477L..45F}) 
the northern source is the dominant source. We speculate that there is extended emission that
contributed to the single-dish observation by \cite{2005ApJS..161..361K}, which is
resolved out by the interferometer. Thus, the fluxes seen by the interferometer
are smaller than expected from the single-dish observations, 
and one would expect that the effect is more pronounced for
the larger core MM3.

MM2 shows no counterpart at 850{\um} or in the FIR, but it does have a counterpart at 3.6\,cm \citep{2002ApJ...570..758M}; 
therefore, the continuum emission from core MM2 is likely dominated by free-free emission from an H{\sc
  ii}-region with only a fraction of the emission from warm dust.

\input{masses}

%% file: masses.tex
\subsection{Estimating the Core Mass}
We present three approaches to roughly estimate the core masses.  The first
method is the LTE mass calculated from the integrated flux of the CS($2-1$)
line transition.  The second method is calculating the virial mass from the
line width of CS spectra.  The third method is based on the 2.7 mm continuum
assuming dust emission and a dust-to-gas ratio. 

\subsubsection{LTE Mass}
Following the standard rotation temperature - column density analysis \citep[e.g.,][]{1991ApJS...76..617T}, 
CS($2-1$) spectra can be used to derive the column density, and further the mass of the
dense gas.  Because the transition CS($2-1$) is not optically thin, a
correction factor needs to be applied to the LTE mass derived from the
assumption of an optically thin line.  For an optically thin molecular line, by
assuming all the transitions are excited by a single temperature
T$_{\text{ex}}$ (LTE) and negligible background continuum emission, the column
density can be calculated from
\citep[e.g.,][]{1995ApJ...445L..59M,1995ApJ...454..782M}

\[
\begin{split}
{\text{N}} \ (cm^{-2}) = & 2.04 \times 10^{20} \times [\frac{1}{\theta_{a} \theta_{b} (arcsec^{2})}] \\ 
                         &\times [\frac{Q e^{E_{u}(K)/T_{ex}}}{\nu^{3}(GHz^3) S \mu^{2}(debye^{2}) g_{k} g_{I}}] \\
                         &\times [\int I(Jy \ beam^{-1}) dv(km \ s^{-1} )] \ , 
\end{split}
\]
where $\theta_{a}$ and $\theta_{b}$ are the FWHM of the synthesized beam, Q is
the partition function, E$_{\text{u}}$ is the upper energy level of the
transition, $\nu$ is the line rest frequency, S is the line strength, and
$\mu^{2}$ is the dipole moment, $g_{k}$ and $g_{I}$ are the usual K-level and
reduced nuclear spin degeneracies.  For our observation,
$\theta_{a}$=5.9$\arcsec$, $\theta_{b}$=4.3$\arcsec$, Q=0.86T$_{\text{ex}}$
\citep{2000tra..book.....R}, E$_{\text{u}}$=7.05 K, $\nu$=97.980968 GHz, S=2,
$\mu$=1.96 debye, and $g_{I}$=$g_{k}$=1. Here we assume an excitation
temperature of 20 K for dense cores forming massive stars \citep[see review
by][]{2007ARA&A..45..481Z}. In the case of an optically thick molecular line,
a correction factor C$_{\tau}$ should be applied
\citep[e.g.,][]{1999ApJ...517..209G}:

$$ \text{N}_{\text{total}} = C_{\tau} \text{N} \ \text{and} $$
$$ C_{\tau}=\frac{\tau}{1-e^{-\tau}} \ , $$
where $\tau$ is the optical depth and can be derived from
$$ \tau = -\ln [1-\frac{T_{MB}}{J(T_{\text{ex}})-J(2.73)}] $$ \citep{2000tra..book.....R},
where T$_{\text{MB}}$ is the main beam brightness temperature and
$$ J(T)=\frac{h\mu}{k} \frac{1}{e^{h\mu /kT}-1}.  $$
Therefore, by assuming uniform densities, the LTE mass can be derived from
$$ M=\mu m_{H} \frac{N}{X} (1.133abD^{2})$$
where $\mu=2.33$ is the mean molecular weight, $m_{H}$ is hydrogen mass, D is
the distance (1.8 kpc for I05345), X is the abundance ratio of CS to H$_{2}$,
which is assumed to be 10$^{-9}$ here \citep{2000tra..book.....R}, and $a$ and
$b$ are the major axis and minor axis of a defined core.
Furthermore, we calculated the number density of molecular hydrogen for these
cores by assuming ellipsoidal cores and $a, \ b$ and $b$ for the three axes:
n$_{H_{2}}$ = $M_{LTE}/2M_{H}/(4\pi /3 (a/2)(b/2)^{2})$.  The column density
and number density of H$_{2}$ range from $2.7\times 10^{22}$ to $11.6\times
10^{22}$ cm$^{-2}$, and from $0.6\times 10^{6}$ to $1.7\times 10^{6}$
cm$^{-3}$, respectively.  The LTE masses of the different cores range from
$\sim$ 2 M$_{\sun}$ to 15 M$_{\sun}$. Table 3 lists the integrated line
intensity ($I_{t} = \int I dv$), the derived
optical depth $\tau$, the column density of CS (N$_{CS}$), 
the number density of H$_{2}$ and the results of LTE mass.

The smallest core mass that our observations are sensitive to is $\sim$
1\,M$_{\sun}$. This is the result of the same calculation as above assuming a
flux equivalent to $3\sigma$ in integrated flux map.
\citet{2009A&A...499..233F} observed their condensations N and S in N$_2$D$^+$
and N$_2$H$^+$ and expected them to be low-mass cores ($<1\,M_{\sun}$)
according to the observed deuterium fraction. Thus, it is not surprising that
we don't detect their condensations N and S.

In the calculation of the LTE masses, a value of 10$^{-9}$ for the CS(2-1) molecular abundance 
is assumed and the abundance is also assumed to be a constant.  However, due to C-bearing molecules 
frozen onto dust grains at early stages of star formation, a central depletion of CS(2-1) is usually seen 
in prestellar cores \citep[e.g.][]{2002ApJ...569..815T,2010MNRAS.407.2434S}, and therefore the 
assumption of a constant abundance may lead to larger uncertainties in the calculation.  Also, according to 
different models or calculations, the derived values for CS(2-1) abundance are various \citep[e.g.][]{2002A&A...393..927B,
2007A&A...461..523P}.  In most literatures, the abundance profile is described by a central depletion and a nearly 
constant value between 10$^{-9}$ and 10$^{-8}$ follows at outer parts of cores.  
Since the radii where the abundance drops is usually very close to the center of prestellar cores, 
we simplify the LTE mass estimation by only adopting the plateau part of the molecular abundance.  
The choice of 10$^{-9}$ sets the upper limit for the calculation and the derived 
LTE masses could vary by about one order of magnitude depending on the chosen values for the abundance.

\begin{table}
\caption{Fitting Parameters For Cores}
\label{tbl:fits}
\begin{tabular}{cccc}
\hline
Label & Amplitude & Peak Position & FWHM \\
      & Jy/beam & km/s & km/s \\
\hline
Core 1 & $0.704\pm 0.002 $ & $-16.592\pm 0.004 $ & $2.626\pm 0.009$ \\
Core 2 & $2.008\pm 0.380 $ & $-18.065\pm 0.004 $ & $1.987\pm 0.053$ \\
        & $-2.040\pm 0.379 $ & $-18.071\pm 0.004 $ & $1.480\pm 0.045$ \\
Core 3 & $1.749\pm 0.032 $ & $-17.659\pm 0.005 $ & $3.237\pm 0.018$ \\
        & $-1.524\pm 0.031 $ & $-17.441\pm 0.003 $ & $1.735\pm 0.016$ \\
Core 4 & $0.699\pm 0.003 $ & $-19.894\pm 0.006 $ & $2.983\pm 0.015$ \\
Core 5 & $0.337\pm 0.002 $ & $-19.172\pm 0.012 $ & $3.940\pm 0.029$ \\
Core 6 & $0.922\pm 0.047 $ & $-17.832\pm 0.021 $ & $0.849\pm 0.049$ \\
Core 7 & $0.921\pm 0.003 $ & $-17.600\pm 0.003 $ & $1.584\pm 0.007$ \\
\hline
\end{tabular}
\end{table}

\subsubsection{Virial Mass}
By assuming the cores are in virial equilibrium, we can derive the virial
mass using the 
line width of CS.  
We assume a standard isothermal density profile for
the cores from \citet{1977ApJ...214..488S}, $\rho = \rho_{0} r^{-2}$
\citep[cf.][]{2003ApJ...592..255L}.  In this case, the virial mass can be
derived from
$$ M_{vir} \sim \frac{3R}{G} \frac{\Delta V^{2}}{8 \ln 2} \ ,$$
where $\Delta V$ is the FWHM of the observed molecular line (see Table 2).
The derived virial mass for each core is listed in Table 3.  
For core 2 and 3 with the double-peak spectrum, we use the line width of the 
emission line to derive the virial masses.  


\subsubsection{Total Mass Estimation from Continuum Emission}
Here we adopt the standard technique to estimate the total mass 
from the continuum emission, in our case the 2.7\,mm continuum data for
core 1 and 3 (remember that core 2 could have considerable free-free
emission; Sect.~\ref{sec:MMobs}).
This method assumes that the continuum emission is from dust and a dust-to-gas ratio.  
By assuming optically thin emission and a single temperature (isothermal),
the total mass (dust + gas) is
$$ M=\frac{F_{\nu} D^{2} R}{B_{\nu}(T_{dust}) \kappa_{\nu}} $$
where F$_{\nu}$ is the flux density, D is the distance to the source, R is the
gas-to-dust ratio, B$_{\nu}(T_{dust})$ is the Planck function at dust
temperature T$_{dust}$, and $\kappa_{\nu}$ is the dust opacity.  The flux
density of core 1 is $22.7 \pm 7.2$ mJy and that of core 3 is $13.8\pm
3.6$ mJy.  For the dust opacity, assuming a gas density $\sim$ 10$^{6}$
cm$^{\text{-3}}$, we adopt the extrapolated value of 0.29 cm$^{2}$ g$^{-1}$ at
$\lambda$=2.7 mm for prestellar core dust with thin ice mantles from
\citet{1994A&A...291..943O}. The extrapolation was done with a power law
($\kappa\propto \lambda^{-\beta}$) and a value of $\beta=1.77$, obtained by
fitting the respective $\kappa$-values in the FIR ($\lambda>300\micron$).
 Therefore, with the assumption of 20 K for the
dust temperature and 100 for the gas-to-dust ratio, we derive the mass of
$18\pm 6\,M_{\sun}$ for core 1 and $11\pm 3\,M_{\sun}$ for core 3.
The derived masses depend much on the assumed values for  the dust
temperature, the opacity, and the dust-to-gas ratio.  The error of the mass
estimates given above only 
reflects the uncertainty in the flux measurement. Taking the uncertainties of
the assumptions into account the estimate is good within one order of
magnitude. For example, if we would assume a rather high temperature of 47\,K
(Paper II)
for core 3 the core mass would only be
$7\,M_{\sun}$. If we would use the opacity for dust grains without ice mantels
the mass would also drop to $7\,M_{\sun}$.

The calculation shows that the total mass from the LTE method and continuum emission of core 1 and core 3 are 
(5$\pm$1 M$_{\sun}$, 18$\pm$6 M$_{\sun}$) and (15$\pm$7 M$_{\sun}$, 11$\pm$3 M$_{\sun}$), respectively.  
Both methods have uncertainties.  For example, 
the results depend on the assumptions of physical conditions and quantities, such as the excitation and 
dust temperature, wavelength dependence of the dust opacity etc.  In CS(2-1) emission, core 3 is 
brighter than core 1, while it is opposite in the 2.7 mm continuum emission 
(see section~\ref{sec:CSobs} and ~\ref{sec:MMobs}).  
This result causes the larger LTE mass of core 3 and larger mass by assuming a dust-to-gas ratio of core 1.  
Other mid-infrared to FIR observations (see paper II) show a brighter core 3 
than core 1, consistent with our CS(2-1) observation.  Therefore, in the following discussion, 
we use the LTE mass to compare it with the virial mass, a method consistent with \citet{2006PASJ...58..343S}.

%% file: discussionI.tex
\section{DISCUSSION}

\subsection{A Collapsing Core}
\label{ssec:cs}

\begin{figure*}
\begin{center}
  \includegraphics[scale=0.5,angle=-90]{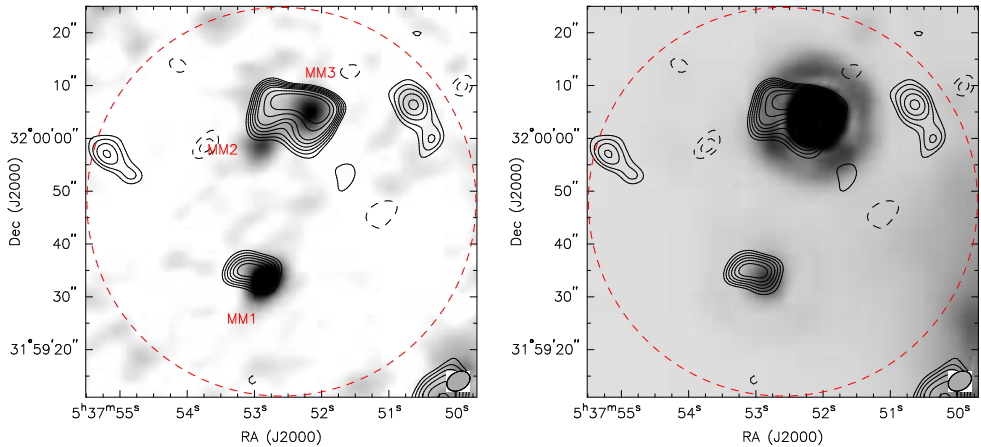}
  \caption{Left panel: CS($2-1$) emission overlaid on the gray-scale 2.7 mm
    continuum data.  The contours represent the CS integrated intensity with the
    same contour levels with Fig. 1.  Right panel: CS($2-1$) emission overlaid
    on the gray-scale Spitzer 24 $\micron$ data.  The contours are the same as
    in the left panel.}
  \label{fig:cont}
\end{center}
\end{figure*}

\begin{figure*}
\begin{center}
   \includegraphics[scale=0.5,angle=-90]{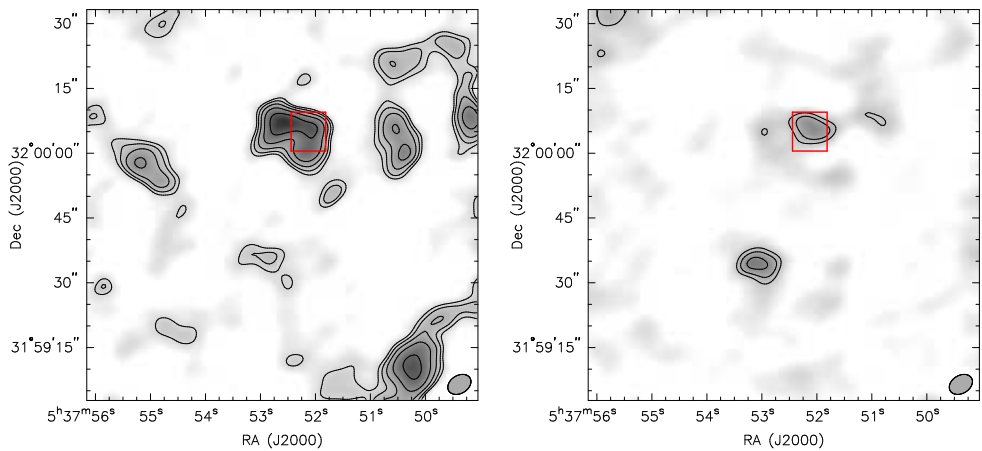}
   \caption{Maps of CS ($2-1$) integrated intensity of the blue
  and red part of the self-absorbed line of core 3.  Left panel: The map of the
  blue emission peak.  The integrated velocities range from -21.6 km s$^{-1}$
  to -17.5 km s$^{-1}$.  The contour levels start from 5$\sigma$ ($\sigma=0.12$
  Jy beam$^{-1}$) and end at 28.3$\sigma$ with a logarithmic step of log($\sqrt{2}\sigma$).  Right
  panel: The map of the right emission peak.  The integrated velocities range
  from -17.1 km s$^{-1}$ to -14.9 km s$^{-1}$.  The contour levels also start
  from 5$\sigma$ ($\sigma$=0.18 Jy beam$^{-1}$) and end at 28.3 $\sigma$ with a
  logarithmic step of log($\sqrt{2}\sigma$). The negative contours are not shown for simplicity.}
   \label{cores}
\end{center}
\end{figure*}

Figure \ref{cores} shows two CS($2-1$) integrated intensity maps integrated
over velocities of the blue and red components of the self-absorption spectrum
of core 3.  The boxes drawn in these two
maps are the same box as in Fig~\ref{fig:csmap}.  Both of the integrated
intensities from the blue and red peak are in the same box.
%

Arguably, this result
suggests that the double-peak structure comes from the same core at the limit
of our resolution (5$\arcsec$),  
although we cannot eliminate the possibility of two cores along the line of sight.  
However, the Gaussian fits in Table \ref{tbl:fits} (emission + absorption) describe the line feature of core 3 considerably well 
and provide an explanation to the double-peaked spectrum as coming from one
core and showing self-absorption.  
Moreover, the absorption is close to the systemic velocity
\citep{2009A&A...499..233F} strengthening the evidence for self-absorption.
However, it is important to note that the negative components in the map are not causing
the self-absorption dip as the negative components seen in Fig~\ref{fig:csmap} are not relevant
in the box of core 3.
To be more specific, we carefully examined the channel maps and there are only few negative contours in the channel where 
the deepest dip occurs (in the velocity of $\sim$ -17 km/s).  
Also, we mapped with short-spacing data removed and only very few negative contours remained, and the deep dip is still seen.  
As a result, the negative contours cannot be the main cause of the dip.  




The double-peaked spectrum of core 3 shows one of the typical characteristics
of gas infall motions: an optically thick transition (CS($2-1$) in our case)
produces a blue-shifted, asymmetrical profile and central dip, with the
blue-shifted line brighter than the red-shifted line
\citep[e.g.,][]{1993ApJ...404..232Z,1995ApJ...448..742C}.  This double-peaked
feature is due to a temperature gradient and velocity gradient between each layer in a collapsing
core (e.g., $v(r)\sim r^{-0.5}$ for the inside-out model) with a static
envelope producing the central self-absorption dip.  The blue-shifted line
comes from the back side of the core and the red-shifted line from the front
of the core.  The lines with higher critical density and higher excitation
temperature near the center will be obscured by the nearby lines with lower
critical density and excitation temperature in the red peak, resulting in a
stronger blue peak than red peak \citep{1999ARA&A..37..311E}.  In this case of I05345,
optically thin lines peak in the dip of the self-absorbed lines.
The absorption peaks approximately at the systemic velocity, also verified by optically
thin tracers measured by \citet{2009A&A...499..233F}.


Similar double-peaked profiles have been observed in CS($2-1$) for different
objects \citep[e.g.,][]{1995ApJ...448..742C,1998ApJ...504..900T,2002A&A...393..927B}.
However, the separation between the blue peak and the red peak in our
observation is significantly larger ($\sim $ 3 km/s) than the separation in
these studies ($\la$ 1 km/s).  This broad separation may arise from the larger
infall velocity for intermediate mass to massive protostars \citep[e.g.][]{2008ApJ...680..446X}.
The absorption feature in the envelope may also affect the separation of the two peaks.
In addition, similar
double-peaked profiles can be produced by a rotating core
\citep{2008ApJ...689..335P}.  Detailed fitting with radiative transfer models
is necessary to better understand other dynamics effects (turbulence, rotation
etc.) in double-peaked profiles \citep{2002A&A...393..927B}.


Although core 2 also shows two peaks in the spectrum, the nature of core 2 is less understood.  
The asymmetry in the spectrum of the two peaks is not significant, indicating that the gas is less likely to be infalling.
The dip is close to the systemic velocity, suggesting that the double-peaked feature may be caused by the self-absorption, 
while the fidelity of core 2 may be affected by the interferometric effect.  


\subsection{Dynamics of Cores}
\label{sec:dynamics}

\begin{table*}
\centering
\begin{minipage}{140mm}
\caption{Physical Properties of CS Cores}
\begin{tabular}{@{}rcrrcrrrr@{}}
\hline
Label & I$_{t}$ & $\tau$ & N$_{CS}$ & n$_{H_{2}}$ & $\Sigma_{H_{2}}$ & M$_{LTE}$ & M$_{vir}$ & P$_{ex}$/k \\
      & (Jy/beam\,km/s)  &          & ($\times$10$^{13}$\,cm$^{-2}$) & ($\times 10^{6}$\,cm$^{-3}$) & (g\,cm$^{-2}$) & (M$_{\sun}$) & (M$_{\sun}$) & (K\,cm$^{-3}$) \\
\hline
Core 1 & $1.94\pm 0.01$ & $0.24 $ & $ 6.12\pm 0.03$ &  1.28$\pm 0.47$ & $0.20\pm 0.00$ & $ 5.3\pm 1.2$ & $26.8\pm  3.4$ & 2.28$\times 10^{8}$ \\
Core 2 & $1.02\pm 1.00$ & $0.21 $ & $ 3.17\pm 3.11$ &  0.59$\pm 0.53$ & $0.11\pm 0.10$ & $ 2.0\pm 2.3$ & $13.2\pm  3.4$ & 0.84$\times 10^{8}$ \\
Core 3 & $3.30\pm 0.13$ & $0.46 $ & $11.55\pm 0.46$ &  1.69$\pm 1.11$ & $0.39\pm 0.02$ & $14.8\pm 6.8$ & $49.8\pm 11.5$ & 4.97$\times 10^{8}$ \\
Core 4 & $2.19\pm 0.01$ & $0.27 $ & $ 7.01\pm 0.03$ &  1.22$\pm 0.81$ & $0.23\pm 0.00$ & $ 7.0\pm 3.5$ & $37.1\pm  9.7$ & 3.24$\times 10^{8}$ \\
Core 5 & $1.39\pm 0.01$ & $0.20 $ & $ 4.29\pm 0.03$ &  0.74$\pm 0.40$ & $0.14\pm 0.00$ & $ 4.7\pm 1.6$ & $67.9\pm 11.4$ & 3.46$\times 10^{8}$ \\
Core 6 & $0.82\pm 0.06$ & $0.36 $ & $ 2.73\pm 0.20$ &  0.61$\pm 0.70$ & $0.09\pm 0.01$ & $ 2.2\pm 2.0$ & $ 2.7\pm  1.3$ & 0.06$\times 10^{8}$ \\
Core 7 & $1.53\pm 0.01$ & $0.38 $ & $ 5.15\pm 0.03$ &  1.04$\pm 0.26$ & $0.17\pm 0.00$ & $ 4.4\pm 0.7$ & $9.6\pm   0.6$ & 0.56$\times 10^{8}$ \\
\hline
\label{tbl:properties}
\end{tabular}
\end{minipage}
\end{table*}

Compared with the line widths for cores in other low-mass star forming regions \citep[e.g.,][]{2002ApJ...575..950O}, 
we observe relatively large line widths from 1.5 km/s to 3.9 km/s for these seven cores.  
The isothermal sound speed for
a region with a temperature 20 K is $\sim$ 0.27 km/s, much smaller than the observed 
line widths.
Therefore, other non-thermal motions play a large role in the
observed broad line widths. There are two possibilities for this non-thermal
line-broadening \citep{2006PASJ...58..343S}.  One possibility is that these
cores formed from rather turbulent gas and still are turbulent.  The
other explanation is that the lines are broadened by 
interactions with the outflow or stellar wind from central protostars.
	
If initial large internal motions played a role in forming cores, higher density is
necessary for a core to bind the system with gravitational energy.  
\citet{2006PASJ...58..343S} suggested such a possible correlation between 
line widths and the average H$_{2}$ density: the larger the line width, the higher the average
density.  Here we did not take the overall line widths of core 2 and 3 into consideration since 
their line widths may be affected and broadened by infall motions as discussed above.  
We see a weak trend (density vs.\ line widths) between core 1, 6 and 7.   
The line widths and average H$_{2}$ density of core 1, 6 and 7 are
$2.63\pm 0.01$, $0.85\pm 0.05$, $1.58\pm 0.01$ km/s and $(1.28\pm 0.47)\times
10^{6}$, $(0.61\pm 0.70) \times 10^{6}$, $(1.04\pm 0.26) \times 10^{6}$
cm$^{-3}$, respectively.  However, considering the large uncertainties in the data, 
the positive trend is less obvious and it is not clear about the role of the initial 
turbulence to the formation of the cores in this region from our data.


Core 4 and 5 do not follow the relation between the averaged density and line widths.  
\citet{2009A&A...499..233F} detected an outflow oriented in the west-east 
direction possibly driven by the 226 GHz continuum source C1-b or C1-a, which corresponds to our core 3.  
The same paper also indicated that this outflow is probably interacting with the
southern portion of the highly deuterated condensation N, which corresponds to
our core 4, and causes the observed broad lines.  Therefore, the line width of core 4 
is more possibly influenced by the interaction with the outflow. 
The red-wing emission of core 4 and the blue-wing emission of core 6 may result from the red and 
blue lobes of the outflow.  
Nevertheless, it is less clear about the mechanism for core 5 to produce the observed spectrum.

However, it seems that the gravitational energy is not enough to bind the
systems.  The derived virial masses are noticeably larger than the LTE masses
except for core 6 (see Table \ref{tbl:properties}).
\citet{2008ApJ...673..315W} studied low to intermediate mass 
cores around MWC 1080 with CS(2-1) and obtained similar LTE masses (calculated from the same method as 
described in Section 3.3.1) and virial masses.
\citet{2006PASJ...58..343S} also concluded that non-turbulent cores have a
similar virial mass to LTE mass but that the virial masses are usually larger than
the LTE masses for turbulent cores.  The line width of core 6 is similar to
that of non-turbulent cores ($\sim$ 0.90 km/s average) and the similarity
between its virial and LTE mass indicates that this core is bound by
gravitational energy.  On the other hand, for
cores with a larger virial mass than LTE mass, external pressure must be
applied to maintain a bound core. The virial equilibrium with
an external pressure (by neglecting magnetic fields and rotation) can be expressed as:
$$
0 = 2U + \Omega - 4\pi R^{3} P_{ex} \ ,
$$  
where U=$\frac{1}{2} M \sigma^{2}$ ($\sigma$: velocity dispersion) is the
kinetic energy and $\Omega=-\frac{3}{5} \frac{G M^{2}}{R}$ is the
gravitational energy.  The results of P$_{ex}$ is listed in Table \ref{tbl:properties}.  Except
core 6, which is already gravitationally bound, other cores need an external
pressure P$_{ex}$/k $\sim$ $10^{8}$ K cm$^{-3}$ to help maintain the structure
of the cores.  With the choice of higher molecular abundance by one order of magnitude, 
the external pressure needed to maintain the cores would decrease to $\sim$ $10^{7}$ K cm$^{-3}$.  
Observations have shown that such pressure is observed in several 
massive star-forming regions \citep[e.g.,][and references therein]{2003ApJ...585..850M}.  Therefore, we suggest that these turbulent cores
with larger kinetic energy may still be 
bound by the external pressure. 



\subsection{Implication for Massive Star Formation}

We suggest that core 3 and core 1 are young stellar objects. They show many indicators 
of high-mass star formation.  
Since young stellar objects are surrounded
by dust and gas, they are not able to be revealed by optical observations (and
shorter wavelength observations) due to dust extinction.  Consequently,
YSOs are often identified by their IR emission excess.  The detection in the Spitzer
24 $\micron$ band is often an indicator of YSOs \citep[e.g.,][]{2008ApJ...678..200C}.
We observed an infrared counterpart of core 3 and 1 in the Spitzer image (Paper II), implying that
both cores contain warm dust, suggestive of a YSO candidate.
In addition, the CS spectrum of core 3 shows the typical signature of infalling gas
motion --- a double-peak feature with the stronger blue-shifted emission than
the red-shifted emission.  The wide separation of the two peaks ($\sim$
3\,km/s) is probably due to the higher masses.

Gas infall is also suggested by fitting the core's SED with a radiative-transfer
model in the companion paper (Paper II).  The radiative-transfer
modeling indicates that  cores 3  and 1 could
be an accreting stellar source  with the central stellar mass of
$\sim5\,M_{\sun}$ surrounded by an envelope with a mass of order of
$10\,M_{\sun}$ (mass estimates in this paper for core 3: $11\,M_{\sun}$ (dust mass),
$17\,M_{\sun}$ (LTE mass) 
and core 1: 4 M$_{\sun}$ (LTE mass), 18 M$_{\sun}$ (dust mass)). 

Although, the virial masses (20 and 50 M$_{\sun}$ for core 1 and 3, respectively) are large compared to 
the LTE masses (4 and 20 M$_{\sun}$), indicating that the cores may be gravitationally unbound, the
cores  can still be bound through the support of the external pressure.  
Therefore, core 3 and 1 are possibly accreting the envelope to form a star
and potentially becoming a massive star ($M_*> 8\,M_{\sun}$).

According to the classification of evolutionary stages of individual high-mass
stars in \citet{2007prpl.conf..165B}, the formation of massive stars start
with High-Mass Starless Cores (HMSCs) and then form High-Mass Protostellar
Objects (HMPOs) via harboring or accreting low/intermediate-mass protostars.
HMPOs are accreting high-mass protostars with masses larger than 8 M$_{\sun}$,
and consist of Hot Molecular Core (HMC), Hypercompact HII regions (HCHIIs,
size $\le$ 0.01 pc) and Ultracompact HII regions (UCHII regions, size $\le$
0.1 pc) at their early phases.  Given the sizes of core 1 and core 3 (0.06
pc $\sim$ 0.08 pc) and the fact that the spectra of core 1 and 3 still peak
in the far-infrared and have not approached near-infrared regimes, we suggest that
these two cores are possibly on their way to become HMPOs.

Also, our cores are consistent with the massive star formation model proposed by
\citet{2003ApJ...585..850M}.  The theory states that turbulent cores in
massive star-forming regions with pressure P$_{ex}$ $\simeq$ $10^{8} \sim
10^{9}$ K cm$^{-3}$ can be gravitationally bound and form stars with high
accretion rates ($10^{-3}$ M$_{\sun}$ yr$^{-1}$) in a short time scale
(several times the free-fall time).  Our cores appear to be turbulent and
non-thermal with line widths ranging from 1.5 km/s to 3.9 km/s.  
However, while the positive correlation between the line widths and average H$_{2}$ density 
implies that the initial turbulence may also influence the formation of these cores,
such correlation is not obvious in our data considering the uncertainties.  
For core 3 and 4, the broad line widths may be 
also contributed by the outflow driven by core 3.  With the exception of core 6, 
the cores have a larger virial mass than LTE mass.  With the help of
external pressure in the parent cloud, these turbulent cores can be
bound and have the potential to become seeds for collapse in the future.



%% file: conclusion.tex
\section{SUMMARY} 

Identifying protocluster members and understanding their dynamics are the first
step to study the initial conditions preserved in the parental clouds and
further unveil the process of star formation.  In this paper, we present 
observations of the intermediate/high mass star forming region IRAS
05345+3157.  The observations have been performed in the line transition
CS($2-1$) with CARMA D array.  At this line frequency, we also observed the continuum
emission at $\lambda=2.7$ mm.  
With our observation, our main conclusions are as follows.

\begin{enumerate}
\item There are seven CS cores identified in the CS moment map.  Both core 1
  and core 3 have counterparts in the 2.7 mm continuum data (MM1 and MM3). 

\item The LTE mass of all the cores are calculated based on the CS(2-1) data.  
The LTE masses range from $\sim$ 2 M$_{\sun}$ to $\sim$ 15 M$_{\sun}$ for all the cores.  
The LTE mass for core 1 is 5$\pm1$ M$_{\sun}$ and that for core 3 is 15$\pm7$ M$_{\sun}$.  
In addition, dust masses are estimated from the continuum data: core 1 has 
the dust mass of $18\pm 6$ M$_{\sun}$ and core 3 has the dust mass of 
$11\pm 3$ M$_{\sun}$.

\item Most of the spectra of the seven cores show a single peak.  
  Core 3 shows a double-peaked spectrum with the blue emission 
  stronger than the red emission, suggesting infall motion of gas.

\item 
Core 1 and 3 are suggested to be intermediate- to high-mass protostellar candidates.  


   
\item The linewidths for the cores are larger than the thermal linewidth at 20 K.
  The broad linewidth of core 3 and 4 are probably contributed by the
  outflow driven by core 3 \citep{2009A&A...499..233F}.  
  However, the role of the initial turbulence to the formation of the cores is not 
  clear by examining the correlation between the linewidths and average H$_{2}$ density with the uncertainties.  
  
\item These cores require an external pressure of
  $\sim \ 10^{8}$ K cm$^{-3}$ to keep them bound.  Such high
  pressure is common among massive-star forming regions
  \citep[e.g.,][]{2003ApJ...585..850M}, suggesting that 
  these cores are possible seeds for future star formation.

\end{enumerate}


%% file: I05345I.bbl
\begin{thebibliography}{}

\bibitem[\protect\citeauthoryear{{Bate} \& {Bonnell}}{{Bate} \&
  {Bonnell}}{2005}]{2005MNRAS.356.1201B}
{Bate} M.~R.,  {Bonnell} I.~A.,  2005, \mnras, 356, 1201

\bibitem[\protect\citeauthoryear{{Belloche}, {Andr{\'e}}, {Despois} \&
  {Blinder}}{{Belloche} et~al.}{2002}]{2002A&A...393..927B}
{Belloche} A.,  {Andr{\'e}} P.,  {Despois} D.,    {Blinder} S.,  2002, \aap,
  393, 927

\bibitem[\protect\citeauthoryear{{Beuther}, {Churchwell}, {McKee} \&
  {Tan}}{{Beuther} et~al.}{2007}]{2007prpl.conf..165B}
{Beuther} H.,  {Churchwell} E.~B.,  {McKee} C.~F.,    {Tan} J.~C.,  2007, in
  {Reipurth} B.,  {Jewitt} D.,   {Keil} K.,  eds, Protostars and Planets V {The
  Formation of Massive Stars}.
pp 165--180

\bibitem[\protect\citeauthoryear{{Bonnell}, {Bate}, {Clarke} \&
  {Pringle}}{{Bonnell} et~al.}{2001}]{2001MNRAS.323..785B}
{Bonnell} I.~A.,  {Bate} M.~R.,  {Clarke} C.~J.,    {Pringle} J.~E.,  2001,
  \mnras, 323, 785

\bibitem[\protect\citeauthoryear{{Bonnell}, {Vine} \& {Bate}}{{Bonnell}
  et~al.}{2004}]{2004MNRAS.349..735B}
{Bonnell} I.~A.,  {Vine} S.~G.,    {Bate} M.~R.,  2004, \mnras, 349, 735

\bibitem[\protect\citeauthoryear{{Caulet}, {Gruendl} \& {Chu}}{{Caulet}
  et~al.}{2008}]{2008ApJ...678..200C}
{Caulet} A.,  {Gruendl} R.~A.,    {Chu} Y.-H.,  2008, \apj, 678, 200

\bibitem[\protect\citeauthoryear{{Choi}, {Evans} II, {Gregersen} \&
  {Wang}}{{Choi} et~al.}{1995}]{1995ApJ...448..742C}
{Choi} M.,  {Evans} II N.~J.,  {Gregersen} E.~M.,    {Wang} Y.,  1995, \apj,
  448, 742

\bibitem[\protect\citeauthoryear{{Evans}
  II}{{Evans}}{1999}]{1999ARA&A..37..311E}
{Evans} II N.~J.,  1999, \araa, 37, 311

\bibitem[\protect\citeauthoryear{{Fontani}, {Caselli}, {Bourke}, {Cesaroni} \&
  {Brand}}{{Fontani} et~al.}{2008}]{2008A&A...477L..45F}
{Fontani} F.,  {Caselli} P.,  {Bourke} T.~L.,  {Cesaroni} R.,    {Brand} J.,
  2008, \aap, 477, L45

\bibitem[\protect\citeauthoryear{{Fontani}, {Zhang}, {Caselli} \&
  {Bourke}}{{Fontani} et~al.}{2009}]{2009A&A...499..233F}
{Fontani} F.,  {Zhang} Q.,  {Caselli} P.,    {Bourke} T.~L.,  2009, \aap, 499,
  233

\bibitem[\protect\citeauthoryear{{Goldsmith} \& {Langer}}{{Goldsmith} \&
  {Langer}}{1999}]{1999ApJ...517..209G}
{Goldsmith} P.~F.,  {Langer} W.~D.,  1999, \apj, 517, 209

\bibitem[\protect\citeauthoryear{{Henning}, {Cesaroni}, {Walmsley} \&
  {Pfau}}{{Henning} et~al.}{1992}]{1992A&AS...93..525H}
{Henning} T.,  {Cesaroni} R.,  {Walmsley} M.,    {Pfau} W.,  1992, \aaps, 93,
  525

\bibitem[\protect\citeauthoryear{{Klein}, {Lee}, {Looney} \& {Wang}}{{Klein}
  et~al.}{2011}]{klein2011}
{Klein} R.,  {Lee} K.~I.,  {Looney} L.~W.,    {Wang} S.,  2011, {Massive
  Star-formation around IRAS\,05345+3157\\II: The Protostars and their
  environment}, in preparation

\bibitem[\protect\citeauthoryear{{Klein}, {Posselt}, {Schreyer}, {Forbrich} \&
  {Henning}}{{Klein} et~al.}{2005}]{2005ApJS..161..361K}
{Klein} R.,  {Posselt} B.,  {Schreyer} K.,  {Forbrich} J.,    {Henning} T.,
  2005, \apjs, 161, 361

\bibitem[\protect\citeauthoryear{{Krumholz}, {Klein} \& {McKee}}{{Krumholz}
  et~al.}{2005}]{2005IAUS..227..231K}
{Krumholz} M.~R.,  {Klein} R.~I.,    {McKee} C.~F.,  2005, in {Cesaroni} R.,
  {Felli} M.,  {Churchwell} E.,   {Walmsley} M.,  eds, Massive Star Birth: A
  Crossroads of Astrophysics Vol.~227 of IAU Symposium, {Radiation pressure in
  massive star formation}.
pp 231--236

\bibitem[\protect\citeauthoryear{{Krumholz}, {Klein} \& {McKee}}{{Krumholz}
  et~al.}{2007}]{2007ApJ...656..959K}
{Krumholz} M.~R.,  {Klein} R.~I.,    {McKee} C.~F.,  2007, \apj, 656, 959

\bibitem[\protect\citeauthoryear{{Krumholz}, {McKee} \& {Klein}}{{Krumholz}
  et~al.}{2005}]{2005ApJ...618L..33K}
{Krumholz} M.~R.,  {McKee} C.~F.,    {Klein} R.~I.,  2005, \apjl, 618, L33

\bibitem[\protect\citeauthoryear{{Lada} \& {Lada}}{{Lada} \&
  {Lada}}{2003}]{2003ARA&A..41...57L}
{Lada} C.~J.,  {Lada} E.~A.,  2003, \araa, 41, 57

\bibitem[\protect\citeauthoryear{{Looney}, {Mundy} \& {Welch}}{{Looney}
  et~al.}{2000}]{2000ApJ...529..477L}
{Looney} L.~W.,  {Mundy} L.~G.,    {Welch} W.~J.,  2000, \apj, 529, 477

\bibitem[\protect\citeauthoryear{{Looney}, {Mundy} \& {Welch}}{{Looney}
  et~al.}{2003}]{2003ApJ...592..255L}
{Looney} L.~W.,  {Mundy} L.~G.,    {Welch} W.~J.,  2003, \apj, 592, 255

\bibitem[\protect\citeauthoryear{{McKee} \& {Tan}}{{McKee} \&
  {Tan}}{2003}]{2003ApJ...585..850M}
{McKee} C.~F.,  {Tan} J.~C.,  2003, \apj, 585, 850

\bibitem[\protect\citeauthoryear{{Mehringer}}{{Mehringer}}{1995}]{1995ApJ...45%
4..782M}
{Mehringer} D.~M.,  1995, \apj, 454, 782

\bibitem[\protect\citeauthoryear{{Miao}, {Mehringer}, {Kuan} \&
  {Snyder}}{{Miao} et~al.}{1995}]{1995ApJ...445L..59M}
{Miao} Y.,  {Mehringer} D.~M.,  {Kuan} Y.-J.,    {Snyder} L.~E.,  1995, \apjl,
  445, L59

\bibitem[\protect\citeauthoryear{{Molinari}, {Testi}, {Rodr{\'{\i}}guez} \&
  {Zhang}}{{Molinari} et~al.}{2002}]{2002ApJ...570..758M}
{Molinari} S.,  {Testi} L.,  {Rodr{\'{\i}}guez} L.~F.,    {Zhang} Q.,  2002,
  \apj, 570, 758

\bibitem[\protect\citeauthoryear{{Onishi}, {Mizuno}, {Kawamura}, {Tachihara} \&
  {Fukui}}{{Onishi} et~al.}{2002}]{2002ApJ...575..950O}
{Onishi} T.,  {Mizuno} A.,  {Kawamura} A.,  {Tachihara} K.,    {Fukui} Y.,
  2002, \apj, 575, 950

\bibitem[\protect\citeauthoryear{{Ossenkopf} \& {Henning}}{{Ossenkopf} \&
  {Henning}}{1994}]{1994A&A...291..943O}
{Ossenkopf} V.,  {Henning} T.,  1994, \aap, 291, 943

\bibitem[\protect\citeauthoryear{{Pavlyuchenkov}, {Wiebe}, {Shustov},
  {Henning}, {Launhardt} \& {Semenov}}{{Pavlyuchenkov}
  et~al.}{2008}]{2008ApJ...689..335P}
{Pavlyuchenkov} Y.,  {Wiebe} D.,  {Shustov} B.,  {Henning} T.,  {Launhardt} R.,
     {Semenov} D.,  2008, \apj, 689, 335

\bibitem[\protect\citeauthoryear{{Pirogov}, {Zinchenko}, {Caselli} \&
  {Johansson}}{{Pirogov} et~al.}{2007}]{2007A&A...461..523P}
{Pirogov} L.,  {Zinchenko} I.,  {Caselli} P.,    {Johansson} L.~E.~B.,  2007,
  \aap, 461, 523

\bibitem[\protect\citeauthoryear{{Rohlfs} \& {Wilson}}{{Rohlfs} \&
  {Wilson}}{2000}]{2000tra..book.....R}
{Rohlfs} K.,  {Wilson} T.~L.,  2000, {Tools of radio astronomy}

\bibitem[\protect\citeauthoryear{{Saito}, {Saito}, {Moriguchi} \&
  {Fukui}}{{Saito} et~al.}{2006}]{2006PASJ...58..343S}
{Saito} H.,  {Saito} M.,  {Moriguchi} Y.,    {Fukui} Y.,  2006, \pasj, 58, 343

\bibitem[\protect\citeauthoryear{{Sault}, {Teuben} \& {Wright}}{{Sault}
  et~al.}{1995}]{1995ASPC...77..433S}
{Sault} R.~J.,  {Teuben} P.~J.,    {Wright} M.~C.~H.,  1995, in {Shaw} R.~A.,
  {Payne} H.~E.,   {Hayes} J.~J.~E.,  eds, Astronomical Data Analysis Software
  and Systems IV Vol.~77 of Astronomical Society of the Pacific Conference
  Series, {A Retrospective View of MIRIAD}.
pp 433--+

\bibitem[\protect\citeauthoryear{{Schreyer}, {Henning}, {Koempe} \&
  {Harjunpaeae}}{{Schreyer} et~al.}{1996}]{1996A&A...306..267S}
{Schreyer} K.,  {Henning} T.,  {Koempe} C.,    {Harjunpaeae} P.,  1996, \aap,
  306, 267

\bibitem[\protect\citeauthoryear{{Seale}, {Looney}, {Chu}, {Gruendl}, {Brandl},
  {Rosie Chen}, {Brandner} \& {Blake}}{{Seale}
  et~al.}{2009}]{2009ApJ...699..150S}
{Seale} J.~P.,  {Looney} L.~W.,  {Chu} Y.-H.,  {Gruendl} R.~A.,  {Brandl} B.,
  {Rosie Chen} C.-H.,  {Brandner} W.,    {Blake} G.~A.,  2009, \apj, 699, 150

\bibitem[\protect\citeauthoryear{{Shu}}{{Shu}}{1977}]{1977ApJ...214..488S}
{Shu} F.~H.,  1977, \apj, 214, 488

\bibitem[\protect\citeauthoryear{{Stahler} \& {Yen}}{{Stahler} \&
  {Yen}}{2010}]{2010MNRAS.407.2434S}
{Stahler} S.~W.,  {Yen} J.~J.,  2010, \mnras, 407, 2434

\bibitem[\protect\citeauthoryear{{Tafalla}, {Mardones}, {Myers}, {Caselli},
  {Bachiller} \& {Benson}}{{Tafalla} et~al.}{1998}]{1998ApJ...504..900T}
{Tafalla} M.,  {Mardones} D.,  {Myers} P.~C.,  {Caselli} P.,  {Bachiller} R.,
   {Benson} P.~J.,  1998, \apj, 504, 900

\bibitem[\protect\citeauthoryear{{Tafalla}, {Myers}, {Caselli}, {Walmsley} \&
  {Comito}}{{Tafalla} et~al.}{2002}]{2002ApJ...569..815T}
{Tafalla} M.,  {Myers} P.~C.,  {Caselli} P.,  {Walmsley} C.~M.,    {Comito} C.,
   2002, \apj, 569, 815

\bibitem[\protect\citeauthoryear{{Turner}}{{Turner}}{1991}]{1991ApJS...76..617%
T}
{Turner} B.~E.,  1991, \apjs, 76, 617

\bibitem[\protect\citeauthoryear{{Wang}, {Looney}, {Brandner} \&
  {Close}}{{Wang} et~al.}{2008}]{2008ApJ...673..315W}
{Wang} S.,  {Looney} L.~W.,  {Brandner} W.,    {Close} L.~M.,  2008, \apj, 673,
  315

\bibitem[\protect\citeauthoryear{{Xue} \& {Wu}}{{Xue} \&
  {Wu}}{2008}]{2008ApJ...680..446X}
{Xue} R.,  {Wu} Y.,  2008, \apj, 680, 446

\bibitem[\protect\citeauthoryear{{Zhang}, {Hunter}, {Brand}, {Sridharan},
  {Cesaroni}, {Molinari}, {Wang} \& {Kramer}}{{Zhang}
  et~al.}{2005}]{2005ApJ...625..864Z}
{Zhang} Q.,  {Hunter} T.~R.,  {Brand} J.,  {Sridharan} T.~K.,  {Cesaroni} R.,
  {Molinari} S.,  {Wang} J.,    {Kramer} M.,  2005, \apj, 625, 864

\bibitem[\protect\citeauthoryear{{Zhou}, {Evans} II, {Koempe} \&
  {Walmsley}}{{Zhou} et~al.}{1993}]{1993ApJ...404..232Z}
{Zhou} S.,  {Evans} II N.~J.,  {Koempe} C.,    {Walmsley} C.~M.,  1993, \apj,
  404, 232

\bibitem[\protect\citeauthoryear{{Zinnecker} \& {Yorke}}{{Zinnecker} \&
  {Yorke}}{2007}]{2007ARA&A..45..481Z}
{Zinnecker} H.,  {Yorke} H.~W.,  2007, \araa, 45, 481

\end{thebibliography}
